
\input phyzzx
\overfullrule=0pt

\date{September 1992}
\Pubnum{\vbox{\hbox{G\"oteborg-ITP-92-40}
	      \hbox{hep-th/9209107}
	      \hbox{\phantom{m}}
	     }}

\titlepage
\title{Quaternionic Superconformal Field Theory}
\author{Martin Cederwall and Christian R. Preitschopf}
\address{Institute of Theoretical Physics,
Chalmers University of Technology, S-412 96, G\"oteborg, Sweden}

\abstract
We develop a superfield formalism
for N=4 superconformal two-dimensional field theory.
A list is presented of minimal free superfields, i.e. of
multiplets containing four bosons and four fermions.
We show that the super-Poincar\'e algebra of the
six-dimensional superstring in the light-cone gauge
is essentially equivalent to a local N=4
superconformal symmetry, and outline the construction
of octonionic superconformal field theory.

\endpage

\REF\kappasiegel{W. Siegel, {\sl Phys.Lett.} {\bf 128B} (1983), 397;
{\sl Class.Quant.Grav.} {\bf 2} (1985) 195.}

\REF\Greenschwarz{
M.B. Green and J.H. Schwarz\journal Phys.Lett.&136B (84) 367.}

\REF\Casalbuoni{R. Casalbuoni\journal Nuovo
Cim.&33A (76) 389.}

\REF\Dirac{P.A.M. Dirac, Lectures on quantum
mechanics, Belfer Graduate School of Science, Yeshiva Univ.
New York (1964).}

\REF\IBMC{I. Bengtsson and M. Cederwall, ITP
G\"oteborg preprint 84-21 (1984).}

\REF\particle{
E. Nissimov, S. Pacheva
and S. Solomon\journal Nucl.Phys.&B297 (88) 349;\nextline
Y. Eisenberg and S. Solomon\journal Nucl.Phys.&B309 (88) 709;\nextline
S.J. Gates Jr et. al.\journal Phys.Lett.&225B (89) 44;\nextline
M.B. Green and C. Hull\journal Phys.Lett.&225B (89) 57;\nextline
R. Kallosh\journal Phys.Lett&224B (89) 273;\nextline
F. Bastianelli, G.W. Delius and E.Laenen\journal Phys.Lett.&229B (89) 223.}

\REF\BBCL{
A.K.H. Bengtsson, I. Bengtsson, M. Cederwall and N. Linden,
\nextline\indent
{\sl Phys.Rev.} {\bf D36} (1987) 1766.}

\REF\IBMCii{I. Bengtsson and M. Cederwall\journal Nucl.Phys.&B302 (88) 81.}

\REF\Berkpart{N. Berkovits\journal Phys.Lett.&247B (90) 45.}

\REF\octonions{M. Cederwall\journal J.Math.Phys.&33 (92) 388.}

\REF\STVZ{
D.P.Sorokin, V.I. Tkach, D.V. Volkov and A.A. Zheltukhin, \nextline\indent
{\sl Phys.Lett.} {\bf 216B} (89) 302.}

\REF\galsok{
A. Galperin and E. Sokatchev, (1992),
Phys. Inst. Univ. Bonn preprint\nextline\indent
BONN-HE-92-07,hep-th/9203051.}

\REF\twistor{R.Penrose and M.A.H. McCallum\journal
Phys.Rep.&6C (72) 241;\nextline A. Ferber\journal
Nucl.Phys.&B132 (78) 55;\nextline T. Shirafuji\journal
Progr.Theor.Phys.&70 (83) 18.}

\REF\stringtwistors{W.T. Shaw\journal Class.Quantum
Grav.&3 (86) 753;\nextline P. Budinich\journal
Comm.Math.Phys.&107 (86) 455;\nextline M. Cederwall\journal
Phys.Lett.&226B (89) 45.}

\REF\Berkovits{N. Berkovits\journal Nucl.Phys.&B358 (91) 169.}

\REF\RNS{P. Ramond,
{\sl Phys. Rev.} {\bf D3} (1971) 2415;\nextline
A. Neveu and J. H. Schwarz,
{\sl Nucl. Phys.} {\bf B31} (1971) 86;\nextline
F. Gliozzi, J. Scherk, and D. Olive,
{\sl Phys. Lett.} {\bf 65B} (1976) 282;\nextline\indent
{\sl Nucl. Phys.} {\bf B122} (1977) 253.}

\REF\spacesheetsusy{
P. Candelas, G.T. Horowitz, A. Strominger and
E. Witten,\nextline\indent {\sl Nucl.Phys.} {\bf B258} (1985) 46;
\nextline
C. Hull and E. Witten, {\sl Phys.Lett.} {\bf 160B} (1985) 398.}

\REF\banksandco{
T. Banks, L.J.  Dixon, D. Friedan and E. Martinec,\nextline\indent
{\sl Nucl.Phys.} {\bf B299} (1988) 613;
\nextline T. Banks and L.J. Dixon,
{\sl Nucl.Phys.} {\bf B307} (1988) 93.}

\REF\DGHS{
F. Delduc, A. Galperin, P. Howe and E. Sokatchev,
Phys. Inst. Univ. Bonn\nextline\indent preprint BONN-HE-92-19,
hep-th/9207050}

\REF\Metali{M. Ademollo et. al.\journal Phys.Lett.&B62 (76)
105;\nextline\indent \sl Nucl.Phys. \bf B111 \rm(1976) 77.}

\REF\Metaliii{M. Ademollo et. al.\journal Nucl.Phys.&B114 (76) 297.}

\REF\extendedsc{the literature on extended superconformal
field theory is enormous. A classification of unitary
representations of the algebras we are interested in
can be found in:\nextline
M. Yu, {\sl Phys.Lett.} {\bf 196B} (1987) 345;\nextline
T. Eguchi and A. Taormina,
{\sl Phys.Lett.} {\bf 200B} (1987) 315.}

\REF\superfields{
K. Schoutens, {\sl Nucl.Phys.} {\bf B295} (1988) 634.}

\REF\uemats{S. Matsuda and T. Uematsu, {\sl Phys.Lett.} {\bf 220B} (1989)
413;\nextline\indent {\sl Mod.Phys.Lett.} {\bf A5} (1989) 841.}

\REF\foursusyintwo{for a review see:\nextline
N.J. Hitchin, A. Karlhede, U. Lindstr\"om and M Rocek,\nextline\indent
{\sl Comm.Math.Phys.} {\bf 108} (1987) 535.}

\REF\KT{T. Kugo and P. Townsend\journal Nucl.Phys.&B221 (83) 357.}

\REF\Sudbery{A.Sudbery\journal J.Phys.&A17 (84) 939.}

\REF\IB{I. Bengtsson\journal Class.Quantum Grav.&4 (87) 1143.}

\REF\FairMan{D.B. Fairlie and C.A. Manogue\journal Phys.Rev.&D36 (87) 475.}

\REF\stringpar{C.A. Manogue and A. Sudbery\journal Phys.Rev.&D40 (89) 4073.}

\REF\confstring{
F. Delduc, E. Ivanov and E. Sokatchev,
Phys. Inst. Univ. Bonn preprint \nextline\indent BONN-HE-92-11
(1992), hep-th/9204071}

\REF\ESTPS{F. Englert, A. Sevrin, W. Troost, A. Van
Proyen and Ph. Spindel,\nextline\indent {\sl J. Math. Phys} {\bf 29} (1988)
281.}

\REF\DTH{F. Defever, W. Troost and
Z. Hasiewicz,\nextline\indent
{\sl Class.Quant.Grav.} {\bf 8} (1991) 253; 257.}

\REF\hypersc{see for example refs.\thinspace\ESTPS, \DTH\ and:\nextline
P.Goddard, W. Nahm, D.I. Olive, H. Ruegg and A. Schwimmer,\nextline\indent
{\sl Commun.Math.Phys.} {\bf 112} (1987) 385.}

\REF\Schafer{R.D. Schafer, An introduction to
nonassociative algebras, New York (1964).}

\REF\BLN{L. Brink, O. Lindgren and B.E.W. Nilsson
\journal Nucl.Phys.&B212 (83) 401.}

\REF\dewitetal{B. de Wit and H. Nicolai
\journal Nucl.Phys.&B231 (84) 506;\nextline F. G\"ursey and
C.H. Tse \journal Phys.Lett.&127B (83) 191;\nextline L.
Castellani and N.P. Warner\journal Phys.Lett.&130B (83)
47;\nextline S. Fubini and H. Nicolai \journal Phys.Lett.&155B
(85) 369.}

\REF\rooman{M. Rooman, {\sl Nucl.Phys.} {\bf B238} (1984) 501.}

\REF\FLBow{E.S. Fradkin and V.Ya. Linetsky, ICTP Trieste
preprint IC/91/348 (1991);\nextline P. Bowcock, Enrico Fermi
Inst. Chicago preprint EFI 92-09 (1992),\nextline\indent
hep-th/9202061.}

\chapter{Introduction}

There is a disturbing lack of understanding of string theory
concerning the underlying gauge principles. This is especially
true of superstring models, where a great obstacle to a
covariant quantum theory is the presence of second-class
constraints following from the so-called kappa-symmetry [\kappasiegel]
for the the covariant Green-Schwarz
action [\Greenschwarz]. The same is true for the
superparticle [\Casalbuoni]. Following the Dirac
procedure [\Dirac], the second class constraints cannot
be eliminated in a covariant way so to yield a covariant
quantum theory [\IBMC]. In the superparticle case one now
knows ways to circumvent this problem [\particle -\STVZ], most of
which involve some kind of twistor formulation
[\twistor,\BBCL -\octonions].

There have been several approaches to string theory within
a twistor framework [\stringtwistors] (the term twistor is somewhat
inappropriate due to the lack of space-time conformal
invariance), but it was only
recently that a twistorial superstring theory appeared
[\Berkovits]. Unfortunately it is again
plagued by second-class constraints,
which now appear in the bosonic sector.
There is however another
interesting feature -- the presence of an extended
superconformal symmetry, which tends to put
rather strong constraints on string models.
Clearly one should develop an appropriate
superfield formalism and make that
symmetry manifest. This is the main objective
of this paper. We will restrict ourselves here to the
case N=4, which pertains to the D=6 superstring.
The structures we describe will
carry over to N=8 and D=10.

The fact that the superstring can be treated
in terms of the RNS model [\RNS] implies
a connection between spacetime and worldsheet supersymmetry.
The relation is more intimate than that,
since one can show that spacetime supersymmetry implies
extended world-sheet supersymmetry [\spacesheetsusy,\banksandco].
There is however no explanation of why it should be so,
presumably because the proofs were given in formulations where
spacetime supersymmetry is not manifest.
We will provide an explanation
in the framework of the superstring
on the light-cone: the super-Poincar\'e algebra is essentially
equivalent to an extended superconformal algebra.

The manifestly supersymmetric string should hence be
considered in a world\-sheet-super\-sym\-metric setting.
The idea of unifying the spacetime-super\-sym\-metric and the
world\-sheet-super\-sym\-metric incarnations of the superstring has
occured to many people in the field.
The idea became a program with the realization that for
certain examples of the superparticle
and superstring with sufficiently extended
worldsheet supersymmetry one can identify the
kappa-symmetry as part of the conformal superdiffeomorphisms
on the worldline or worldsheet [\STVZ,\galsok,\DGHS]. This could mean that
one can avoid second-class constraints by gauge-fixing
that kind of action in the proper, yet unknown, fashion.
We may then envisage a spacetime covariant superconformal
gauge that features a finite number of free fields
and first class constraints implemented \'a la BRST.
If such a gauge does not exist then
the new superstring formulations will probably not be
terribly attractive. If it does we will obtain
some kind of extended superconformal field theory.

Extended superconformal algebras were first constructed in
refs.\thinspace\Metali, \Metaliii\ and have by now been studied in some detail
[\extendedsc]. Typically those algebras are much
larger than the ones we are interested in,
which contain the minimal number of
generators. In addition, we are aiming at a formalism that allows
at least the computation of the superstring S-matrix, and
for that purpose it is surely convenient to have a
linear algebra realized on a small number of free fields,
with as much symmetry as possible manifest.
Neither existing superfield formalisms [\superfields,\uemats] nor
earlier studies of representations of N=4 supersymmetry
in the context of hyperk\"ahler sigma-models [\foursusyintwo]
are therefore tailormade for our purposes.

We believe that the division algebra formalism is a natural
tool for the superconformal symmetries occurring in string theory.
It it well known [\KT,\Sudbery] that there exists a close relationship
between the division algebras {\bf R}, {\bf C}, {\bf H} and {\bf O}
and N=1 supersymmetry in D=3,4,6,10 or, by dimensional
reduction, N=1,2,4,8 supersymmetry in D=2.
The reason is that those hypercomplex numbers very efficiently
encode the important Gamma-matrix identity
$$
\gamma_{a(b}^\mu \; \gamma_{cd)}^\mu\; = 0 \ .
\eqno\eq
$$
Consequently they have been used in the analysis of the
Green-Schwarz superstring [\IB, \FairMan, \IBMCii, \octonions, \stringpar]
and twistor(-like) superstring models [\Berkovits,\confstring],
and of course also in the description of extended
two-dimensional superconformal algebras [\hypersc].
We will make extensive use of quaternions and intend this
to be a stepping stone towards octonions.

The paper is organized as follows: in section 2 we briefly discuss
the quaternion language we will use. A translation table to
$\gamma$-matrix notation is given in the Appendix.
In section 3 we describe all free superconformal fields
with minimal field content, \ie\ with 4 bosonic and
4 fermionic components, and describe their energy-momentum
superfields and correlation functions. The list is smaller than
expected, since one can show that minimal ghost systems with arbitrary
conformal weight do not exist. In section 4 we present
actions for all the fields.
Subsequently we apply our technology to the Lorentz-algebra
of the light-cone superstring in D=6. Our construction
does not make reference to the coordinates of the internal
space, but only to the superconformal generators.
We conclude with an outline of the N=8 construction.

\chapter{Quaternionic formalism}

We will often use a quaternionic language in this
paper. This allows us to dispense with complicated
index structures and simplifies some calculations
and results. It is also important conceptually,
since various transformations can be clearly separated,
and this makes the step from N=4 to N=8 almost
straightforward.

In the Appendix we provide a translation table from
quaternion notation to $\sigma$-matrix notation
that we will frequently use in the rest of the paper.

Our conventions are as follows. For an element $h=\sum_{n=0}^3
h_a e_a \in \bf H$ (the quaternionic ring), we define {\it
conjugation} by
$$
 h\rightarrow h^* =h_0-\sum_{i=1}^3h_ie_i \;
. \eqno\eq
$$
 It is distinguished from complex conjugation,
written as $z\rightarrow \overline z$. The complex imaginary
unit is not affected by $\bf H$ conjugation. The {\it real} or
{\it scalar} part of h is
$$
[h]=\coeff{1}{2} (h+h^*)\;
,\eqno\eq
$$
 and the {\it imaginary} or {\it vector} part
$$
\{h\}=\coeff{1}{2}(h-h^*)\;.\eqno\eq
$$
 The notation in the
last two definitions is non-standard but useful. A quaternion
is called a {\it unit} quaternion if $hh^*=1$, i.e. if it has
unit norm.

\section{Quaternionic representations}

In this section, we will describe how the tensor algebra of
some relevant compact Lie groups is expressed in terms of
elements of the division algebra {\bf H} of quaternions.
The results
presented can not be regarded as original
(see e.g. refs.\thinspace \Sudbery, \KT, \IBMCii), but since
our formalism depends so heavily on them, they deserve to be
presented in some detail.

\subsection{$SU(2)$ and $SU(2)^2$}

A {\it spinor} $\lambda\in{\bf H}$ is defined to transform
under $SU(2)$ as
$$
\lambda\rightarrow\lambda
h_1\;,\eqn\rightmult
$$
 where $h_1$ is a unit quaternion,
$h_1h_1^*=1$, i.e. a point on $S^3\approx SU(2)$. It is
obvious that the transformations
$$
\lambda\rightarrow
h_2^*\lambda\; ,\;\; h_2h_2^*=1\;,\eqn\leftmult
$$
parametrize
another $SU(2)$ that commutes with the first one. We are
naturally led to consider the group $SU(2)^2\approx SO(4)$.
We call the spinor representation ``{\bf 4}''.

The reader will probably be used to
complex spinorial representations of $SU(2)$.
The quaternionic spinors (four real components)
can be understood in terms of two-component complex
spinors as follows. The imaginary unit quaternions fulfill the
Clifford algebra of $SU(2)$:
$$
e_ie_j+e_je_i=-2\delta_{ij}\;.\eqn\clifford
$$
They may
therefore be realized as $i$ times the Pauli matrices. Let
$$
{\bf H} \ni\lambda =\lambda_0{\bf 1}
+i\lambda_j\sigma_j\;,\eqno\eq
$$
where $\sigma_i$ are Pauli
matrices. Then
$$
\lambda
=\left(\matrix{\lambda_0+i\lambda_3&i\lambda_1+\lambda_2\cr
i\lambda_1-\lambda_2&\lambda_0-i\lambda_3\cr}\right)
\equiv\left(\matrix{\mu_1&\mu_2\cr}\right)\;,\eqno\eq
$$
and the two two-component spinors $\mu_\alpha$ are related by
$$
\overline\mu_\alpha
=i\sigma_2\varepsilon_{\alpha\beta}\mu_\beta\;.\eqn\majorana
$$
This is what in ref.\thinspace\KT\ is called an $SU(2)$-Majorana
condition. The representation {\bf 4} is a real irreducible
representation of $SU(2)^2$.

Using the Clifford algebra of eq.\clifford , one realizes that
tensor products of representations are encoded in quaternionic
multiplication. If we denote the $SU(2)$'s in eqs. \rightmult\
and \leftmult\ by $SU(2)_1$ and $SU(2)_2$ respectively, it is
clear that multiplication of two spinors
$\lambda,\mu\in{\bf 4}$ yields
$$
\{\lambda^*\mu\}\in{\bf
3}_1\;,\;\;\{\lambda\mu^*\}\in{\bf 3}_2\;,\eqn\vectorrep
$$
with obvious meaning of the subscripts. With $A_1\in{\bf
3}_1$, $A_2\in{\bf 3}_2$ we also have
$$
\lambda A_1\in{\bf
4}\;,\;\;A_2\lambda\in{\bf 4}\;.\eqno\eq
$$
The vector representations in eq. \vectorrep\ cover all six
antisymmetric products in ${\bf 4}\times{\bf 4}$. In real
formalism, we can form projection operators on the two
orthogonal (selfdual and anti-selfdual) subspaces in
\vectorrep :
$$
\lambda_{[a}\mu_{b]}=
(\Pi^{1}_{abcd}+\Pi^{2}_{abcd})\lambda_c\mu_d\;,\eqno\eq
$$
where
$$
\eqalign{\Pi^{1}_{abcd}&=
\coeff{1}{2}(\delta^{ab}_{cd}-\coeff{1}{2}\varepsilon_{abcd})
=-\coeff{1}{4}[e_a^*e_be_{[c}^*e_{d]}]
=\coeff{1}{4}\sigma^M_{a[c}(1)\overline{\sigma}^M_{d]b}(1)
\;,\cr
&=\coeff{1}{4}\sigma^j_{ab}(-1)\sigma^j_{cd}(-1)
=\coeff{1}{16}\sigma^{\mu\nu}_{ab}(-1)\sigma^{\mu\nu}_{cd}(-1)
\cr
\Pi^{2}_{abcd}&=\coeff{1}{2}(\delta^{ab}_{cd}+\coeff{1}{2}\varepsilon_{abcd})
=-\coeff{1}{4}[e_ae^*_be_{[c}e^*_{d]}]
=\coeff{1}{4}\sigma^\mu_{a[c}(-1)\overline{\sigma}^\mu_{d]b}(-1)
\cr
&=\coeff{1}{4}\sigma^J_{ab}(1)\sigma^J_{cd}(1)
=\coeff{1}{16}\sigma^{MN}_{ab}(1)\sigma^{MN}_{cd}(1)
\;.\cr}\eqn\projectop
$$
These projections will play a crucial r\^ole in defining the
superfields.
The various $\sigma$-matrices are defined in the Appendix, and
are to be understood as chiral projections of $\gamma$-matrices.

\subsection{$SU(2)^3$}

The form of the projection operators in eq.\projectop\ appears
``non-covariant'', considering it contains an
$\varepsilon$-symbol in ``spinor'' indices. However, if we
reconsider the transformation rules \rightmult\ and \leftmult
, we find that $\lambda$ is actually what is usually
recognized as a {\it vector} of $SO(4)\approx SU(2)_1\otimes
SU(2)_2$. When we later specify the field content of our
superconformal field theory, we will have bosons and fermions
(four real degrees of freedom each) transforming under
``vector'' and ``spinor'' representations of $SO(4)$ (the
group of euclidean rotations on the six-dimensional light-cone
superstring variables). The interpretation of the $SU(2)^2$
treated above is that $SU(2)_1$ is half the group of space
rotations $SO(4)\approx SU(2)_0\otimes SU(2)_1$, and that
$SU(2)_2$ is an internal group (as in eq. \majorana ).

Considering the symmetry group $SU(2)_0\otimes
SU(2)_1\otimes SU(2)_2$, we have three inequivalent
quaternionic representations, which we denote ${\bf 4}_v$,
${\bf 4}_s$ and ${\bf 4}_c$. The transformation rules are ($L$
and $R$ denote left and right multiplication by a unit
quaternion and $L^*$ and $R^*$ by its conjugate):
$$
\vbox{\halign{\hfil #\hfil&&\quad\hfil #\hfil\cr &${\bf
4}_v$&${\bf 4}_s$&${\bf 4}_c$\cr $SU(2)_0\;\;$&$R$&$L^*$&1\cr
$SU(2)_1\;\;$&$L^*$&1&$R$\cr
$SU(2)_2\;\;$&1&$R$&$L^*$\cr}}\eqn\transformations
$$
The
composition rules are symbolically, with {\bf H}
multiplication,
$$
{\bf 4}_v{\bf 4}_s\rightarrow{\bf
4}_c^*\;\;\hbox{and cyclic permutations,}\eqno\eq
$$
 and
express the triality of $SU(2)^3$, formally identical to the
triality of $SO(8)$ as expressed in terms of octonionic
multiplication  [\Schafer]. This
means that, as for $SO(8)$, the choice of ``vector''
representation is arbitrary, as seen from \transformations .

When $SU(2)^3\rightarrow SU(2)^2$ by only considering group
elements $h_0\in SU(2)_0$ and $h_1\in SU(2)_1$ with $h_0=h_1$,
\transformations\ is reduced to
$$
\vbox{\halign{\hfil
#\hfil&&\quad\hfil #\hfil\cr &${\bf 4}_v$&${\bf 4}_s$&${\bf
4}_c$\cr ${\it diag}(SU(2)_0,SU(2)_1)\;\;$&$L^*R$&$L^*$&$R$\cr
$SU(2)_2$&1&$R$&$L^*$\cr}}\eqno\eq
$$
 so the the
representations split as
$$
\eqalign{{\bf 4}_v&\rightarrow{\bf
1}\oplus{\bf 3}_1\;,\cr {\bf 4}_s&\rightarrow{\bf 4}^*\;,\cr
{\bf 4}_c&\rightarrow{\bf 4}\cr}\eqno\eq
$$
 as
$SU(2)^3\rightarrow SU(2)^2$ (${\bf 4}^*$ is not inequivalent
to {\bf 4} --- this is just to denote that the conjugate
spinor obeys the transformation rules \rightmult\ and
\leftmult ). Observe again the exact formal analogue of the
splitting of the three {\bf 8}'s of $SO(8)$ under
$SO(8)\rightarrow SO(7)$.

\subsection{$SU(2)^4$}

We may extend the transformation table \transformations\
to four $SU(2)$'s:
$$
\eqalign{
v \longrightarrow &\qquad h_1^* \ v \ h_0 \cr
s \longrightarrow &\qquad h_0^* \ s \ h_3 \cr
c^* \longrightarrow &\qquad h_1^* \ c^* \ h_2 \cr
w \longrightarrow &\qquad h_2^* \ w \ h_3 \ ,\cr}\eqno\eq
$$
where $h_A$ are four unit quaternions parametrizing $SU(2)_A$.
\transformations\ is obtained by setting $h_2 = h_3$.
Clearly $SO(4)_1 \approx SU(2)_0 \otimes SU(2)_1$ and
$SO(4)_2 \approx SU(2)_2 \otimes SU(2)_3$
commute. We denote with $v^\mu$ and $w^M$ the corresponding
vector representations.
The $\sigma$-matrices satisfy
$$
\sigma^\mu_{ab}(-1)\sigma^N_{bc}(1)=\sigma^N_{ab}(1)\sigma^\mu_{bc}(-1)\ .
\eqn\commute
$$
Note that spinor indices may carry different
chirality with respect to the two $SO(4)$'s.

Identity \commute\ makes some of the subsequent constructions
possible. For N=8 it is valid only up to
terms proportional to the associator of octonions.

It is difficult to write \commute\ using complex spinors.
That is the reason the $SO(4)$-covariant form of the N=4
algebras we shall present was not written down already in
ref.\thinspace\Metaliii.

\chapter{Superfields}

\section{Scalar Superfields}

Consider a general scalar holomorphic field on the
superspace spanned by the complex coordinate z and the spinor
$\theta =\theta_ae_a$. Its component expansion is
$$
\eqalign{
\Phi(z,\theta)=&A(z)+[\theta^*\!B(z)]-\coeff{1}{2}[\theta^*\!\theta
C^*(z)]
-\coeff{1}{2}[\theta\theta^*\!D^*(z)]+[\theta^{3*}\!E(z)]+\theta^4F(z)
\cr
=&A(z) + \theta_a B_a(z) -
 \coeff{1}{8}\theta_a \sigma^{\mu\nu}_{ab}(-1)\theta_b\ C^{\mu\nu}(z)
-\coeff{1}{8}\theta_a\sigma^{MN}_{ab}(1)\theta_b\ D^{MN}(z)\cr
&+\theta^{3}_a \!E_a(z)+\theta^4F(z)
\;,\cr}
\eqn\phiexp
$$
where $\theta^3=-\coeff{1}{6}\theta\theta^*\!\theta$ (i.e.
$\theta^3_a=-\coeff{1}{6}\varepsilon_{abcd}\theta_b\theta_c\theta_d$)
and
$\theta^4=\coeff{1}{24}\theta^*\!\theta\theta^*\!\theta
=\coeff{1}{24}\varepsilon_{abcd}\theta_a\theta_b\theta_c\theta_d$.
In the following, $z$-dependence will be suppressed when
dealing with holomorphic fields. According to the discussion
in section (3.1), the transformation properties of the
component fields are: $A,F\in {\bf 1}$; $B,E\in {\bf 4}$;
$C\in {\bf 3}_1$ and $D\in {\bf 3}_2$.
In $SO(4)$-language, $C$ and $D$ are antisymmetric selfdual
tensors.
Note that the above interpretation implies a choice of
spinor chiralities with respect to $SO(4)_1$
and $SO(4)_2$. We could just as well have made another
choice by barring the appropriate $\sigma$-matrices in
\phiexp.
We note that in ref.\thinspace\Metaliii\ some
results similar to ours were derived using a complex
formalism.

\subsection{Constraints}

We define the covariant spinor derivative
$$
{\cal
D}=\partder{}{\theta}+\theta\partial\eqn\covariantder
$$
 with
$\partder{}{\theta}=e_a\partder{}{\theta_a}$ and $\partial
=\partder{}{z}$. In order to reduce the field content to four
bosonic and four fermionic fields, we impose the following constraint on
$\Phi$:
$$
\{{\cal DD}^*\}\Phi=0\;,\eqn\dualityconstr
$$
 i.e. we
use the projection operators of eq.\projectop\ and demand
$\Pi^2_{abcd}{\cal D}_c{\cal D}_d\ \Phi=0$. This constraint is quite
analogous to the selfduality condition in N=4 super-Yang-Mills
theory [\BLN]. The component
expansion yields
$$
\eqalign{D&=0\;,\cr E&=\partial B\;,\cr
F&=\partial^2\!A\;.}\eqno\eq
$$
According to eq. \covariantder,
$w(\theta)=-\coeff{1}{2}$ ($w$ denotes conformal weight),
and the constrained, {\it antiselfdual}, superfield
$$
\Phi=\varphi+[\theta^*\psi]-
\coeff{1}{2}[\theta^*\!\theta\chi^*]+
[\theta^{3*}\partial\psi]+\theta^4\partial^2\varphi\eqn\selfdualfield
$$
has by definition $w(\Phi)=0$.
Its $\theta^2$-term is antiselfdual in the spinor indices.
Hence the name.
$\Phi$ contains exactly the components
corresponding to the light-cone superstring in D=6, provided
we interpret the $w=1$ vector as the derivative of a
$w=0$ field, $\chi^j=\partial{\varphi}^j$. This
assignment will be justified in section (4.2).

\subsection{Correlators}

For the moment, we assume that the field $\Phi$ is
``self-conjugate'', and we assign correlators to its
components according to their conformal weights:
$$
\eqalign{&\varphi(z)\varphi(\zeta)\sim \ln (z-\zeta)\;,\cr
&\psi_a(z)\psi_b(\zeta)\sim{\delta_{ab}\over z-\zeta}\;,\cr
&\chi^i(z)\chi^j(\zeta)\sim{\delta_{ij}\over
(z-\zeta)^2}  \;.\cr}\eqn\corrcollection
$$
These are collected into the
supercorrelator
$$
\Phi(Z_1)\Phi(Z_2)\sim\Delta^{(e)}(Z_1,Z_2)
\equiv\ln(z_{12}-[\theta_1^*\theta_2])-e{\theta_{12}^4\over
z_{12}^2}= ln(Z_{12}) -e{\theta_{12}^4\over
Z_{12}^2} \;, \eqn\supercorr
$$
where
$e=1$ for an antiselfdual field,
$Z=(z,\theta)$ and $z_{12}=z_1-z_2$, while
$\theta_{12}$ $=$ $\theta_1^a - \theta_2^a$ and
$Z_{12} = z_1 - z_2 - \theta_1^a \theta_2^a$
are the supersymmetry-invariant distance functions.
The last term ensures that the function
$\Delta^{(+)}$ satisfies the constraint \dualityconstr . For a field
of opposite duality it takes the opposite sign. We will say more about
correlators when examining vector superfields.

\section{The Energy-momentum Superfield}

{}From the components of the selfdual superfield of
eq.\selfdualfield , we may construct the generators of an N=4
superconformal algebra:
$$
\eqalign{&J=\coeff{1}{2}\psi\psi^*\;,\cr
&G=\psi(\partial\varphi-\chi)\;,\cr
&L=\coeff{1}{2}\partial\varphi\partial\varphi-
\coeff{1}{2}[\psi^*\partial\psi]+\coeff{1}{2}\chi^*\chi\;.\cr}
\eqn\lxrep
$$
We have thus a set of $SU(2)_2$ Kac-Moody generators $J^K = \coeff{1}{2}
\sigma^K_{ab}(1)\ \psi_a \psi_b$ $ \in{\bf
3}_2$ $(w=1)$, four fermionic generators $G_a = \sigma^\mu_{ab}(-1)\ \psi_b\
\partial \varphi^\mu$ $\in{\bf 4}$
$(w=\coeff{3}{2})$ and a Virasoro generator $L= \coeff{1}{2}
\partial\varphi^\mu\partial\varphi^\mu - \coeff{1}{2}\psi_a \partial
\psi_a$ $\in{\bf 1}$ $(w=2)$.
As far as the energy-momentum algebra is concerned,
it makes no difference whether we declare that
$(\partial \varphi, \partial \varphi^j)$
transforms as ${\bf 1} + {\bf 3}_1$ or as ${\bf 4}^v_1$.
The Kac-Moody currents transform
as antisymetric selfdual tensors under $SO(4)_2$ and
that is the reason none of the $\varphi^\mu$'s
have to appear in $J$, making its contruction feasible.
The fact that the $SO(4)$'s commute assures the closure of the
algebra.

The operator product expansions are
$$
\eqalign{\cr
&J_u(z)J_v(\zeta)\;\sim\;{-c/3\over(z-\zeta)^2}\; [u^*v]\;+\;{1\over
z-\zeta}J_{[u,v]}\;\;,\cr\cr
&J_u(z)G_\alpha(\zeta)\;\sim\;{1\over
z-\zeta}G_{u\alpha}\;\;,\cr\cr
&G_\alpha(z)G_\beta(\zeta)\;
\sim\;{  2c/3\over(z-\zeta)^3}\;[\alpha\beta^*]\;
+\;{2\over(z-\zeta)^2}J_{\{\alpha\beta^*\}}(\zeta)\;\cr
&\phantom{ G_\alpha(z)G_\beta(\zeta)\;\sim\;   }
+\;{1\over
z-\zeta}(\partial
J_{\{\alpha\beta^*\}}+2[\alpha\beta^*]L)\;\;,\cr\cr
}$$
$$\eqalign{
&L(z)J(\zeta)\;\sim\;{1\over(z-\zeta)^2}J(\zeta)\;+\;{1\over
z-\zeta}\partial J\;\;,\cr\cr
&L(z)G(\zeta)\;\sim\;{3/2\over(z-\zeta)^2}G(\zeta)\;+\;{1\over
z-\zeta}\partial G\;\;,\cr\cr
&L(z)L(\zeta)\;\sim\;{c/2\over(z-\zeta)^4}\;
+\;{2\over(z-\zeta)^2}L(\zeta)\;+\;{1\over
z-\zeta}\partial L\;\;,\cr\cr}\eqn\energycorr
$$
where a
variable $X$ in the quaternionic representation {\bf r} is
given an explicit (bosonic) quaternionic index $\rho\in{\bf
r}$ by the contraction $X_\rho=[\rho^*X]$ and the central terms
are expressed in terms of the conformal anomaly
$c=6$. This is the N=4
superconformal algebra of ref.\thinspace\Metaliii\ in
the notation of ref.\thinspace\ESTPS.

We now want to gather the superconformal generators into a
superfield of a well-defined type. If the superfield is not to
contain inverse powers of $\partial$, it should contain $J$ as
its lowest component, and thus as a whole belong to the
representation ${\bf 3}_2$. It is not difficult to check that
the field
$$
{\cal L}=\coeff{1}{2}{\cal D}\Phi{\cal
D}^*\!\Phi\eqn\lexpansion
$$
 has the right properties. Its component
expansion is
$$
{\cal L}=\{\;J+\theta
G^*+\theta\theta^*(\coeff{1}{2}\partial J+L)-\theta^3\partial
G^*-\theta^4\partial^2\!J\;\}\;.\eqno\eq
$$

Compared to a general superfield in ${\bf 3}_2$, with
$3\cdot 2^4$ components, ${\cal L}$ is highly constrained. The
form of the constraint that projects down to the minimal
superfield shown above is
$$
[{\cal D}\Omega^*]{\cal L}+\{{\cal D}\Omega^*\!{\cal
L}\}=0\;,\eqn\energyconstr
$$
where
$\Omega\in{\bf 4}$ is a constant spinor.
Its real-component version reads
$$
{\cal D}_a{\cal
L}^I-\coeff{1}{2}\varepsilon^{IJK}\sigma^J_{ab}(1){\cal D}_b{\cal
L}^K=0\;,\eqno\eq
$$
where we have used the Appendix.

The action of the stress tensor on a primary superfield
$\Xi_A(z,\theta)$
is given by:
$$
\eqalign{
{\cal L}^J(Z_1)\; \Xi_A(Z_2) \ \sim \ &
\left(
  \delta^{JK} {1\over Z_{12}}
- \coeff{1}{2} \epsilon^{JLK}
 { \theta_{12} \sigma^L(e) \theta_{12}\over Z_{12}^2}\;
- 2 e  \delta^{JK} {\theta_{12}^4\over Z_{12}^3}
\right)\ \delta^K \Xi_A(Z_2)
\cr\cr &
+\left(
   {\theta_{12 a} \over Z_{12}}
+ e  {\theta_{12 a}^3 \over Z_{12}^2}
\right)
\; \sigma^J(e)_{ab} \; {\cal D}_b  \Xi_A(Z_2)
\cr\cr &
+ \; w \; { \theta_{12} \sigma^J(e) \theta_{12}\over Z_{12}^2}\;
\Xi_A(Z_2)
\ +\   {\theta_{12} \sigma^J(e) \theta_{12} \over Z_{12}}
\; \partial \Xi_A(Z_2)
\ .\cr}
\eqn\primfield
$$
Here $e=1$, $w$ denotes the conformal weight of $\Xi_A$ and
$\delta^K \Xi_A$ describes the variation of $\Xi_A$ under an
infinitesimal $SU(2)$-rotation with index $K$, for example:
$\delta^K {\cal D}_a \Phi = -\sigma^K_{ab}(1) {\cal D}_b\Phi$,
$\delta^K \Phi = 0$ and
$\delta^K {\cal L}^L = 2 \epsilon^{KLJ} {\cal L}^J$.
We note that the structure of \primfield\ does not
depend on $\Xi_A$ being a minimal field with only four
bosonic and four fermionic components.
An example is provided by the $SU(2)$-scalar
$\Xi = \exp(ik\Phi)$, which is primary with $w=-\coeff{1}{2}k^2$,
but contains both selfdual and antiselfdual
components at the $\theta^2$-level.

Accordingly,
the superspace rendition of \energycorr\ is
$$
\eqalign{
{\cal L}^J(Z_1)\; {\cal L}^K(Z_2) \ \sim \ &
\ - \ {c\over 3}\;
\left(
\delta^{JK} {1 \over Z_{12}^2}\; +\;
\epsilon^{JKL}\;  { \theta_{12} \sigma^L(e) \theta_{12}\over Z_{12}^3}\;
- \; 6 e \; \delta^{JK} {\theta_{12}^4 \over Z_{12}^4}
\right)
\cr\cr &
+ \left(
  \delta^{JI} {1\over Z_{12}}
- \coeff{1}{2} \epsilon^{JLI}
 { \theta_{12} \sigma^L(e) \theta_{12}\over Z_{12}^2}\;
- 2 e  \delta^{JI} {\theta_{12}^4\over Z_{12}^3}
\right)\ 2 \epsilon^{IKN} {\cal L}^N(Z_2)
\cr\cr &
+\left(
  {\theta_{12 a} \over Z_{12}}
+ e  {\theta_{12 a}^3 \over Z_{12}^2}
\right)
\; \sigma^J(e)_{ab} \; {\cal D}_b  {\cal L}^K(Z_2)
\cr\cr &
+ \; 2 \; { \theta_{12} \sigma^J(e) \theta_{12}\over Z_{12}^2}\;
 {\cal L}^K
\ +\   {\theta_{12} \sigma^J(e) \theta_{12} \over Z_{12}}
\; \partial  {\cal L}^K(Z_2)
\ .\cr}
\eqn\superemtcorr
$$
Again, $e=1$. The other sign choice would be appropriate
for an antiselfdual energy-momentum multiplet.
All the functions of $Z_{12}$ and $\theta_{12}$
that appear in \primfield\ and \superemtcorr\
can be written as derivatives of the basic correlator
\supercorr.

\section{Vector Superfields}

Scalar superfields
cannot describe the D=6 light-cone superstring.
They have only one bosonic zero-mode.
However, a look at the energy-momentum multiplet
reveals that we may interprete the bosonic
components as a vector of $SO(4)_1$.
Then of course the
$SO(4)$ of space rotations
for the six-dimensional superstring is ``broken'' in \selfdualfield\ to
$SU(2)$, and only $SU(2)^2$ of the $SU(2)^3$-symmetry is manifest.

In order to restore the full symmetry
we demand a superfield that transforms as ${\bf
4}_v$ under $SU(2)^3$. It must also contain no other
components than this bosonic ${\bf 4}_v$ and a fermionic ${\bf
4}_s$ (or ${\bf 4}_c$).
We may construct it from $\Phi$, once we observe
that, although in this interpretation
$\Phi$ is non-covariant, the derivative
${\cal D}_a\Phi \equiv \Psi_a$ is:
$$
\Psi = \psi + \theta\partial X + \coeff{1}{2} \theta \theta^*
    \partial \psi    - \theta^3 \partial^2 X - \theta^4 \partial^2 \psi \ .
\eqno\eq
$$
Then the vector superfield $\cal X$ can be obtained from $\Phi$ by
$$
[\Omega^*{\cal D]X}= \Omega^*\Psi
\ .
\eqno\eq
$$
This equation also encodes the constraints:
$$
 {\cal D}_a {\cal X}^\mu = \sigma^\mu_{ab}(-1) \Psi_b
\qquad \iff \qquad
{\cal D}_a {\Psi}_b = \sigma^\mu_{ab}(-1) \partial{\cal X}^\mu
 \ .
\eqn\usex
$$
The equivalence holds of course only up to zero modes of the
differential operators involved.
The equations above are quite useful for superspace
calculations.
Once we set $X\in {\bf 4}_v$, $\psi\in{\bf 4}_s^*$ and
$\theta\in{\bf 4}_c$, the equivalent
constraints
$$
\{{\cal D}^*\Omega{\cal X}^*\}=0\;,\eqn\vectorconstraint
$$
are manifestly $SU(2)^3$-covariant and imply the following
component expansion of the antiselfdual vectorfield:
$$
{\cal
X}=X+\theta^*\psi+\coeff{1}{2}\theta^*\!\theta\partial
X+\theta^{3*}\partial\psi+\theta^4\partial^2\!X\;.\eqno\eq
$$
Eq. \vectorconstraint\ implies that the $\theta^2$-term
is antiselfdual in the spinor indices.
Note that in $\Psi$ the $\theta^2$-term is selfdual.
\vectorconstraint\ also relates the first and third
components of the superfield. By construction we have
$\Phi={\cal X}^0 $ and the energy-momentum superfield
reads as for the antiselfdual scalar
$$
{\cal L} = \coeff{1}{2} \Psi \Psi^*  \ .  \eqno\eq
$$
It is straightforward to write down the correlator for
the $\cal X$ field.
One uses \usex\ to
show ${\cal D}^*{\cal D} \Phi = 4\partial {\cal X}$
and then derives from the $\Phi$-correlator \supercorr\
$$
{\cal X}_u(Z_1){\cal
X}_v(Z_2)\sim[vu^*\underline\Delta^{(e)}(Z_1,Z_2)]\;,\eqn\supervectcorr
$$
where $e=1$ in the antiselfdual case and
$$
\eqalign{ \underline\Delta^{(e)}(Z_1,Z_2)
=&
\ln\left(Z_{12}\right) +
{1\over 2} { \theta_{12}^* \theta_{12} \over Z_{12}}
-{e \over 24} { \theta_{12}^* \theta_{12} \theta_{12}^* \theta_{12}
\over Z_{12}^2}
\cr
=&
\ln\left(Z_{12} + {1\over 2} { \theta_{12}^* \theta_{12}}
+{3-e \over 24}
{ \theta_{12}^* \theta_{12} \theta_{12}^* \theta_{12} \over Z_{12}}
\right)
\;\;.\cr}\eqn\vectorcorr
$$
The quaternionic logarithm is
defined by its series expansion.

Since the scalar field is so closely related to the vector field,
we can use the identities \usex\ also in
that context. The scalar field correlator
\supercorr\ is simply the real part of the vectorcorrelator
\vectorcorr:
$$
\Delta^{(e)}(Z_1,Z_2)=[\underline\Delta^{(e)}(Z_1,Z_2)]\;.\eqno\eq
$$

\section{Ghost Systems}

The superfields $\Phi$ and ${\cal X}$ may be thought of as
matter multiplets of a gauge fixed version of some
``covariant'' theory (the case at hand, with N=4, corresponds
to the six-dimensional superstring). The local superconformal
symmetry may then be a remnant of the gauge symmetry of the
covariant model. In order to treat this presumed gauge
invariance, it is of interest to consider supermultiplets
of ghost fields. It turns out that
their $SO(4)$-index structure restricts
the conformal weight assignments in a nontrivial manner.

\subsection{Reparametrization Ghosts}

We now consider conjugate pairs of ghosts $(B^J,C^K)$,
$(\beta_a,\gamma_b)$ and $(b,c)$ of conformal weights $(\lambda-1,2-\lambda)$,
$(\lambda-\coeff{1}{2},\coeff{3}{2}-\lambda)$ and $(\lambda,1-\lambda)$
respectively, with correlators
$$
\eqalign{
C^I(z)B^J(\zeta)&\sim{\delta^{IJ}\over z-\zeta}\;,\cr
\gamma_a(z)\beta_b(\zeta)\ &\sim{\delta_{ab}\over z-\zeta}\;,\cr
c(z)b(\zeta)\ \ \ &\sim{1\over z-\zeta}\;.\cr}\eqn\bccorrel
$$
For $\lambda=2$ this set of fields corresponds to the
ghost system for super - reparametrizations.

We constructed the most general bilinear expressions
for the currents $J$, $G$ and $L$, and demanded that they satisfy
\energycorr\ for arbitrary central terms. Some algebra shows
that this implies $\lambda=2$ and $c=12$, and up to trivial
rescalings the currents are unique:
$$
\eqalign{
J^K =&\ \beta_a \sigma^K_{ab}(1) \gamma_b \
      -\  2 \epsilon^{KLM} B^L C^M\
      -\  \partial ( B^K c ) \cr
G_a =&\  - \partial \beta_a c \ -\  \coeff{1}{2} \beta_a \partial c
      \ + \ 2 \gamma_a b + \sigma^J_{ab}(1) \beta_b C^J
      \ - \ 2 \sigma^J_{ab}(1) \partial \gamma_b B^J
      \ -\  \sigma^J_{ab}(1) \gamma_b \partial B^J\cr
L\ \    =&\   2 \partial c b\  +\  c \partial b
     \  -\  \coeff{3}{2} \partial \gamma_a \beta_a
     \  -\  \coeff{1}{2} \gamma_a \partial \beta_a
     \  +\  \partial C^J B^J      \ .\cr
}\eqn\ghostrepcurr
$$
By $\oint {dz \over 2\pi i} \;
G_a = \coeff{\partial}{\partial\theta_a} - \theta_a \partial_z$
the above normalization implies the superfield expansions
$$
{\cal C}\ = \ c \ + \ 2\theta_a \gamma_a \ + \ \theta_a \sigma^J_{ab}(1)
            \theta_b C^J \ - \ 2 \theta^3_a \partial \gamma_a \ - \
            \theta^4 \partial^2 c
\eqn\cexpans
$$
appropriate for a $w=-1$ selfdual scalar field obeying the contraint
$$
\{{\cal D^*D\}C}=0\;,\eqn\cconstraints
$$
and
$$
\eqalign{
{\cal B}^J \ =& \ B^J \ + \ \theta_a \sigma^J_{ab}(1) \;\beta_b \ +
             \theta_a \sigma^J_{ab}(1) \theta_b \;b \ - \
             \coeff {1}{2} \theta_a \sigma^{J}(1)_{ab} \sigma^{K}(1)_{bc}
                            \theta_c\;
                   \partial B^K \cr
            & \ - \ \theta^3_a \sigma^J_{ab}(1)\; \partial \beta_b \ -
             \ \theta^4 \partial^2 B^J\ ,
}
\eqn\bexpans
$$
which implies that
$\cal B$ is of the same type as
$\cal L$ and hence carries $w = 0$ and satisfies
$$
\{{\cal D}\Omega^*{\cal B}\}+[{\cal
D}\Omega^*]{\cal B}=0\;.\eqn\bconstraints
$$
In order to obtain the supercorrelator, one forms
the selfdual scalar field
$$
\eqalign{
{\bf B}  =& \ -\coeff{1}{24} \left[ {\cal D} {\cal D}^* {\cal B}^*\right]\cr
         =& \ b \ + \ \coeff{1}{2} \left[ \theta \partial \beta^*\right]
            \ - \  \coeff{1}{4} \left[ \theta \theta^* \partial^2 B^* \right]
            \ - \  \coeff{1}{2} \left[ \theta^3 \partial^2 \beta^* \right]
            \ - \ \theta^4 \partial^2 b \ .\cr}
$$
Using the constraints \bconstraints\  one derives
$\{{\cal D D^*}\} {\bf B}=2\partial^2 {\cal B}$, and from the scalar correlator
$ {\bf B}(Z_1) {\cal C}(Z_2) \sim \partial_1 \Delta^{(+)}(Z_1,Z_2) $
one then gets
$$
{\cal B}(Z_1){\cal C}(Z_2)
\sim{ \{ {\cal D}_1 {\cal D}^*_1 \}\over
2\partial_1}\Delta^{(-)}(Z_1,Z_2)\ .\eqno\eq
$$
The ${1\over\partial}$ is superficial. The superfield identities
that were used in this calculation are, by themselves, tricky.
They are most easily understood by realizing that
$(\partial c, -2 C^J)$ and $2 \gamma_a$ form, as far as the
supersymmetry constraints are concerned, a
(selfdual) vectormultiplet much like the (antiselfdual)
multiplet $\partial \varphi^\mu$ and $\psi_a$
described previously. One then employs the analogs of \usex.
Similar considerations hold for the fields
$(b, \partial B^J/2)$ and $\beta_a / 2$.

The energy-momentum
multiplet \ghostrepcurr\ reads in superfield language
$$
{\cal L}^{gh}
=\left\{
\coeff{1}{4}   \{ {\cal DD^*}  \} {\cal C B}
+\coeff{1}{4}{\cal DCD^*B}
+\partial({\cal CB})
\right\}\;.\eqno\eq
$$
We also display the canonical N=4 super-reparametrization part of the
BRST operator:
$$
\eqalign{
{\cal Q} =&
\oint{dz\over 2\pi i}\ \ \left( c \left(L^\Phi + \coeff{1}{2}L^{gh}\right)
              - \gamma_a \left(G^\Phi_a + \coeff{1}{2}G^{gh}_a\right)
              +  C^K \left(J^\Phi_K + \coeff{1}{2}J^{gh}_K \right)
		\ \right)\cr
=& \oint{dz\over 2\pi i}\ \Big(  c L^\Phi\ -\ \gamma_a G^\Phi_a \ +\
                                 C^K J^\Phi_K
  \cr
 & \phantom{\oint{dz\over 2\pi i}\ \Big( }
               \ + \ 2 \ (\ c\partial c b + c \partial C^K B^K
                         - \epsilon^{IJK} C^I C^J B^K
  \cr
 & \phantom{\oint{dz\over 2\pi i}\ \Big( \ + \ 2 \ ( \ }
                         - c \partial \gamma_a \beta_a - \gamma_a \gamma_a b
                         - \gamma_a \sigma^J_{ab}(1) \beta_b C^J
                         + \gamma_a \sigma^J_{ab}(1) \partial \gamma_b B^J )
		\Big)
  \cr
=&- \coeff{1}{12}\oint{dz\over 2\pi i} \Big[
 \ {\cal D}_a \sigma^J_{ab}(1) {\cal D}_b \ \Big( {\cal C}
 \left( {\cal L}^\Phi_J + \coeff{1}{2}{\cal L}^\Phi_J \right) \Big) \
\cr
& \phantom{- \coeff{1}{12}\oint{dz\over 2\pi i} \Big[ \ }
 + \coeff{1}{2}
 \left( {\cal D}_a \sigma^J_{ab}(1) {\cal D}_b \; {\cal C} \right)
 \ \left( {\cal L}^\Phi_J + \coeff{1}{2}{\cal L}^\Phi_J \right)
\cr
& \phantom{- \coeff{1}{12}\oint{dz\over 2\pi i} \Big[ \ }
 - \coeff{1}{2}\ {\cal C}\ {\cal D}_a \sigma^J_{ab}(1) {\cal D}_b
\left( {\cal L}^\Phi_J + \coeff{1}{2}{\cal L}^\Phi_J \right)
 \Big]_{\theta=0} \cr}
\eqno\eq
$$
The BRST current is an unconstrained scalar superfield of conformal
dimension 1, whose $\theta_a$-component is a total derivative.
${\cal Q}$ is therefore invariant under supersymmetry transformations.
In order to achieve manifest supersymmetry we use
${\bf L}  = \ -\coeff{1}{24} \left[ {\cal D} {\cal D}^* {\cal L}^*\right]$
and write
$$
{\cal Q} = -\coeff{1}{2}
 \oint{dz\over 2\pi i}\; d\theta^4\ \left( {1\over \partial^2}
{\cal C} \right) \; {\bf L}\ ,
\eqno\eq
$$
where the inverse derivatives vanish in the component expansion
after some partial integrations and the fermionic integral is
normalized such that $\int d\theta^4 \; \theta^4 = 1$.

Since $c^\Phi=6$ and $c^{gh}=12$, we cannot hope for
${\cal Q}^2=0$ beyond the classical level. If we choose the
matter sector of our covariant theory to consist of, say, 6
selfdual scalar superfields, we must set constraints that restrict
the physical field content to just one of them.
These contraints might be implemented in BRST-fashion with
yet other ghosts, which then must make the
total conformal anomaly vanish.

The above BRST current
is not a component of a minimal field, and hence
we cannot argue that the superstring, once properly
gauge-fixed to N=4 superconformal gauge, should
be built from a collection of minimal fields because
those are {\it the} natural objects in conformal field
theory. However, their simplicity and economy of size
still makes them attractive as building blocks.

\subsection{Other Minimal Ghosts}

The somewhat surprising result that a minimal ghost system
${\cal C} \in {\bf 1}$ and ${\cal B} \in {\bf 3}$
has a unique conformal weight leads us to consider
further representation assignments.
The only remaining choices with minimal field
content are ${\cal C} \in {\bf 4}$ and ${\cal B} \in {\bf 4}$,
constrained like vector superfields:
$$
\eqalign{
{\cal D}_u {\cal C}_v =& \left[ u^* v \Gamma \right] \cr
{\cal D}_u {\rm B}_w =& \left[ u^*  {\cal B} w \right] \ .\cr
}
\eqn\bcconstr
$$
We denote superfields according to their base components:
the expansion of ${\cal C}$ starts with the fermion $c$, $ \rm B$
has the bosonic base component $\beta$ etc.
\bcconstr\ can be interpreted as
$$
\eqalign{
{\cal D}_a {\cal C}_b = \overline\sigma^\mu_{ab}(-1) \Gamma^\mu
\qquad \iff & \qquad
{\cal D}_a {\Gamma}^\mu = \overline\sigma^\mu_{ab}(-1) \partial{\cal C}_b
\cr
{\cal D}_a {\rm B}^\mu = \overline\sigma^\mu_{ab}(-1) {\cal B}_b
\qquad \iff & \qquad
{\cal D}_a {\cal B}_b = \overline\sigma^\mu_{ab}(-1) \partial{\rm B}^\mu
\ \;\cr}
\eqn\cinspinor
$$
or as
$$
\eqalign{
{\cal D}_a {\cal C}^M = \sigma^M_{ab}(1) \Gamma_b
\qquad \iff & \qquad
{\cal D}_a {\Gamma}_b = \sigma^M_{ab}(1) \partial{\cal C}^M
\cr
{\cal D}_a {\rm B}_b = \sigma^M_{ab}(1) {\cal B}^M
\qquad \iff & \qquad
{\cal D}_a {\cal B}^M = \sigma^M_{ab}(1) \partial{\rm B}_b
\ .\cr}
\eqn\cinvector
$$
Any attempt to assign the fermionic variables $\theta$ to
a vector representation results in breaking $SO(4)$-covariance.
Therefore we do not consider that possibility any further.
We assume free field correlators of the form
\bccorrel, and superconformal generators that
are bilinear in the fields.
The constraints \bcconstr\ then determine the
generators $G_a$ up to a total derivative term:
$$
G = -\partial c \beta + b \gamma + \rho \partial (c\beta)
\eqn\generalg
$$
We may actually read this equation as one in superspace,
after simply replacing $c$ by ${\cal C}$ and so on.
Then \generalg\ implies that the parameter $\rho$ parametrizes the
conformal weights of the ghost multiplet.
We now demand closure of the superconformal algebra
and obtain the result that either $\rho=0$ or $\rho=1$,
which in turn means that the component fields form
a weight-$(1,0)$ and a weight-$(1/2,1/2)$ quadruplet.

We now give a list of the possible minimal vectorghost
systems.
In the following
one may chose $e=\pm 1$
in each case,
and of course also bar or unbar all the
$\sigma$-matrices.
The energy-momentum superfield can read off directly from the
component expressions for the fields $J_{ab}$.
Note that not all of those choices correspond to \bcconstr,
since they require different quaternion products.

\bigskip
\input tables
\thicksize=0pt
\begintable
$c=-12$	| $c_a$& $b_a$	& $\gamma^\mu$	& $\beta^\mu$ 	\cr
w	|   0	&   1	&      1/2	&      1/2
\endtable
\vskip -\bigskipamount
$$
\eqalign{
J_{ab}\  \equiv& \ \sigma^k_{ab}(e) J^k =
	-\overline\sigma^{\mu\nu}_{ab}(e)\ \gamma^\mu \beta^\nu \cr
G_a\  =& \  \overline\sigma^{\mu}_{ac}(e)\ b_c \gamma^\mu
	-\overline\sigma^{\mu}_{ab}(e)\ \partial c_b \beta^\mu \cr
L\ \   =& \ \partial c_a b_a
	- \coeff{1}{2} \partial \gamma^\mu  \beta^\mu
	 + \coeff{1}{2} \gamma^\mu \partial \beta^\mu \ . \cr
}\eqn\vectorghostone
$$

\bigskip
\medskip

\begintable
$c=12$	| $c_a$& $b_a$	& $\gamma^\mu$	& $\beta^\mu$ 	\cr
w	|   1/2	&   1/2	&      1	&      0
\endtable
\vskip -\bigskipamount
$$
\eqalign{
J^K\  =& \ b^a \sigma^K_{ab}(-e)  c^b \cr
G_a\  =& \  \overline\sigma^{\mu}_{ac}(e)\ b_c \gamma^\mu
	+\overline\sigma^{\mu}_{ab}(e)\  c_b \partial\beta^\mu \cr
L \ \   =& \ \coeff{1}{2}\partial c_a b_a
	-\coeff{1}{2} c_a \partial b_a
 	+  \gamma^\mu \partial \beta^\mu \ .
\cr
}\eqn\vectorghosttwo
$$

\bigskip

The constraints \cinspinor\ are appropriate for
\vectorghostone\ and \vectorghosttwo\ with $e=-1$.
Case \vectorghosttwo\ is somewhat trivial, since it corresponds to a pair of
${\cal X}^\mu$-type vector fields.
The supercorrelator can be easily derived using \cinspinor:
$$
\eqalign{
{\cal C}_a(Z_1) {\rm B}^\nu(Z_2)\ \sim& \
\coeff{1}{4} \sigma^\mu_{ab}(e) {\cal D}_{1 b} \underline\Delta^{(-e)\mu\nu}
(Z_1,Z_2) \cr
\ =& \
\coeff{1}{4} \sigma^\mu_{ab}(e) {\cal D}_{1 b}
\left(
\delta^{\mu\nu}\ln\left(Z_{12}\right) +
\coeff{1}{2} { \theta_{12} \overline\sigma^{\mu\nu} \theta_{12} \over Z_{12}}
+ e \delta^{\mu\nu} { \theta_{12}^4 \over Z_{12}^2}
\right)\cr
\ =& \
\sigma^\nu_{ab}(e)\left( {\theta_{12b} \over Z_{12} }
 - e  { \theta_{12b}^3 \over Z_{12}^2 }\right) \ .
\cr}
\eqn\cinspcorr
$$
This is the basic correlation function for the previous two
ghost systems. Any other is computed by taking derivatives
of it.

\bigskip
\medskip

\begintable
$c=-12$	|$c^\mu$&$b^\nu$& $\gamma_a$	& $\beta_a$ 	\cr
w	|   0	&   1	&      1/2	&      1/2
\endtable
\vskip -\bigskipamount
$$
\eqalign{
J^K\  =& \ \beta^a \sigma^K_{ab}(-e)  \gamma^b  \cr
G_a\  =& \  \overline\sigma^{\mu}_{ac}(e)\  \gamma_c b^\mu
	-\overline\sigma^{\mu}_{ab}(e)\ \beta_b  \partial c^\mu \cr
L \ \  =& \  \partial c^\mu b^\mu
	- \coeff{1}{2} \partial \gamma_a  \beta_a
	 + \coeff{1}{2} \gamma_a \partial \beta_a \ .
\cr
}\eqn\vectorghostthree
$$

\bigskip
\medskip

\begintable
$c=12$	|$c^\mu$&$b^\nu$& $\gamma_a$	& $\beta_a$ 	\cr
w	|   1/2	&   1/2	&      1	&      0
\endtable
\vskip -\bigskipamount
$$
\eqalign{
J_{ab}\   \equiv& \ \sigma^k_{ab}(e) J^k =
	\overline\sigma^{\mu\nu}_{ab}(e)\ c^\mu b^\nu \cr
G_a\  =& \  \overline\sigma^{\mu}_{ac}(e)\  \gamma_c b^\mu
	+\overline\sigma^{\mu}_{ab}(e)\ \partial c^\mu \cr
L \ \  =& \  \coeff{1}{2} \partial c^\mu b^\mu
	-  \coeff{1}{2} c^\mu  \partial b^\mu
	+ \gamma_a  \partial \beta_a \ .
\cr
}\eqn\vectorghostfour
$$

The constraints \cinvector\ with barred $\sigma$-matrices
apply for \vectorghostthree\
and \vectorghostfour\ with $e=1$.
Case \vectorghostfour\ is the
triality-rotated version of \vectorghosttwo.
Here the bosons are protected from the
$J$-currents not by being a representation of the other $SO(4)$, but
by their chirality.
The supercorrelator reads:
$$
\eqalign{
{\cal C}^\mu(Z_1) {\rm B}_a(Z_2)\ \sim& \
 \sigma^\mu_{ab}(e)\left( {\theta_{12b} \over Z_{12} }
 + e  {\theta_{12b}^3 \over Z_{12}^2 }\right) \ .
\cr}
\eqn\cinvecorr
$$

\chapter{Actions}

This section deals with the problem of finding proper actions
for the types of fields discussed so far. There is one
apparent problem present, associated with models with N$>$2.
Namely, if one writes an action schematically,
$$
S=\int
d^2\!z\int d(\hbox{fermions}){\cal S}\;,\eqno\eq
$$
 and counts
the scaling weights $s$, one obtains the relation ($s(S)=0$)

$$
0=-2+\coeff{1}{2}n_f+s({\cal S})\;,\eqno\eq
$$
 where $n_f$ is
the number of fermionic variables appearing in the measure.
Since $\cal S$ should be constructed without inverse
derivatives from a scalar field $\Phi$ with $s(\Phi)=0$, there
is a limit on $n_f$:
$$
n_f\leq 4\;.\eqn\fermionbound
$$
We also have to introduce a separate set of
fermionic variables to accomodate the anti-holomorphic fields,
so this means that we must restrict the fermionic integration
from the ``full'' measure to a smaller one.
One then has to check the supersymmetry invariance of
the action separately.

\section{Vector Fields}

In order to describe a left- and a rightmoving
vector field with common zero modes,
we start with an unconstrained field
${\cal X}^\mu(z,\overline z , \theta, \tilde\theta)$.
The first constraint we impose is
$$
{\cal D}_a {\cal X}^\mu = \sigma^\mu_{ab}(-1) \Psi_b \ .
\eqn\offveconleft
$$
The next constraint then has to be
$$
\tilde {\cal D}_a {\cal X}^\mu =
\overline\sigma^\mu_{ab}(-1) \tilde \Psi_b \ .
\eqn\offveconright
$$
This implies that $\theta\in{\bf 4}_c$ and $\tilde{\theta}\in{\bf
4}_s^*$ and we are thus constructing a type IIa superstring.
If one unbars the $\sigma$-matrix in \offveconright\
to obtain left- as well as rightmoving fermions with the
same chirality, the constraint system implies
equations of motion and thus becomes illegal.
If one replaces a $\sigma(-1)$ with
$\sigma(1)$ one obtains equivalent results.
\offveconleft\ and \offveconright\ together imply the
following component expansion:
$$
\vbox{
\halign{\hfil #\hfil&&\quad\hfil#\hfil\cr
${\cal X}$ & &=& $X$ & $+$ & $\theta^*\psi$ & $+$ & $\coeff{1}{2}
\theta^*\!\theta\partial X$ & $+$ & $\ldots$
\cr
&&$-$&$\tilde\psi^*\tilde{\theta}$&$+$&$\theta^*\eta
\tilde{\theta}$&$-$&$\coeff{1}{2}\theta^*\!\theta\partial\tilde\psi^*
\tilde{\theta}$&$+$&$\ldots$\cr
&&$-$&$\coeff{1}{2}\overline{\partial}X\tilde{\theta}^*\!
\tilde{\theta}$&$-$&$\coeff{1}{2}\theta^*\overline{\partial}\psi
\tilde{\theta}^*\!\tilde{\theta}$&$-$&$\coeff{1}{4}\theta^*\!
\theta\partial\overline{\partial}X\tilde{\theta}^*\!
\tilde{\theta}$&$+$&$\ldots$\cr
&&$+$&$\ldots$&&&&&&\cr}}\eqn\vectoroffexp
$$
In this quaternion notation it is immediately
obvious that $(\psi, \eta)$ is a vector multiplet in the
$\tilde\theta$-direction. In order to facilitate
writing the Lagrangian, we note
the action of derivatives on ${\cal X}$:
$$
\eqalign{
{\cal D}_a \Psi_b = \sigma^\mu_{ab}(-1) \partial {\cal X}^\mu
\ \ \ & \ \ \
\tilde{\cal D}_a \tilde\Psi_b =
\overline\sigma^\mu_{ab}(-1) \overline\partial {\cal X}^\mu
\cr
{\cal D}_a \tilde\Psi_b = -\sigma^M_{ab}(1) N^M
\ \ \ & \ \ \
\tilde{\cal D}_a \Psi_b =
\overline\sigma^M_{ab}(1) N^M
\cr
{\cal D}_a N^M = - \sigma^M_{ab}(1) \partial \tilde \Psi_b
\ \ \ & \ \ \
\tilde{\cal D}_a N^M =
\overline\sigma^M_{ab}(1) \overline\partial \Psi_b
\cr
}
\eqno\eq
$$
The $SO(4)_2$-vector superfield $N^M$ contains $\eta^M$ as its base
component. The above equations may also be read
as the supersymmetry transformations of the base components of
the relevant superfields.
It is now easy to show that
$$
\eqalign{
S \ =& \ - \coeff{1}{32} \int d^2z\; d\theta_a d\tilde\theta_b
\left(   {\cal D}_a {\cal X}^\mu   \tilde{\cal D}_b {\cal X}^\mu
\right) \cr
=& \ \coeff{1}{2} \int d^2z \left(
\partial X^\mu \overline\partial X^\mu
\ + \ \eta^M \eta^M \ - \ \psi_a \overline\partial \psi_a
\ - \ \tilde\psi_a \partial \tilde\psi_a \right)
\ \cr
}
\eqn\vectoraction
$$
is off-shell supersymmetric and $SU(2)^4$-invariant.
We leave open the question of a type IIb vector field action
that is manifestly N=4 supersymmetric and
$SO(4)$-invariant.

\section{Scalar Fields}

The constraint system for the scalar field
$\Phi(z,\overline z , \theta, \tilde\theta)$
reads:
$$
{\cal D}_a \sigma^{MN}_{ab}(1) {\cal D}_b \ \Phi = 0
\eqn\scaconone
$$
$$
\tilde {\cal D}_a \sigma^{MN}_{ab}(1) \tilde {\cal D}_b \ \Phi = 0
\eqn\scacontwo
$$
$$
{\cal D}_a \tilde {\cal D}_b \ \Phi =
\coeff{1}{4} \delta_{ab} {\cal D}_c \tilde {\cal D}_c \ \Phi
\ + \ \coeff{1}{16} \sigma^{MN}_{ab}(1)
{\cal D}_c \sigma^{MN}_{cd}(1) \tilde {\cal D}_d \ \Phi \ .
\eqn\scaconthree
$$
Not surprisingly, the solution to these constraints is the
real part of \vectoroffexp.
The components $\chi^j$ and $\tilde\chi^j$ that appear
at level $\theta^2$ and $\tilde\theta^2$
due to \scaconone\ and \scacontwo\ satisfy
$$
\partial \tilde\chi^j - \overline\partial\chi^j = 0
\eqn\hhhchichi
$$
as a consequence of \scaconthree.
Hence we may write
$$ \chi^j = \partial X^j \ \ \ {\rm and} \ \ \
\tilde\chi^j = \overline\partial X^j \ ,
\eqn\hhhchipart
$$
up to zero modes. As long as we only work with $\Phi$
and its derivatives, that ambiguity does not matter.
This is the promised justification of the assignment
$\chi^j=\partial{\varphi}^j$ in section (3.1).

The action is the same as that for the vector superfield,
namely
$$\eqalign{
S \ =& \ - \coeff{1}{32} \int d^2z\; d\theta_a d\tilde\theta_b
\left( \sigma^\mu_{ac}(-1) \Psi_c \overline\sigma^\mu_{bd}
(-1) \tilde\Psi_d
\right)\cr
=& \ - \coeff{1}{32} \int d^2z\; d\theta_a d\tilde\theta_b
\left( \Psi_a \tilde\Psi_b - \coeff{1}{4}
\sigma^{\mu\nu}_{ac}(-1) \Psi_c \sigma^{\mu\nu}_{bd}
(-1) \tilde\Psi_d
\right)
\cr}
\eqn\scalaraction
$$
and hence off-shell supersymmetric.
In the first version of the action
the left- and rightmoving fermions carry opposite
$SO(4)_1$-chirality, in the second one they have the same.
The equality in \scalaraction\ should be understood
component by component.
The components are then interpreted as belonging to
different $SO(4)$-covariant objects.

\section{Heterotic Strings}

For vector fields
${\cal X}^\mu(z,\overline z , \theta)$
we set the constraints \offveconleft, which guarantee
that the action
$$
\eqalign{
S \ =& \ \coeff{1}{8} \int dz^2 \; d\theta_a \;\Big[
 {\cal D}_a {\cal X}^\mu\; \overline\partial {\cal X}^\mu
\Big]
\cr
=& \ \coeff{1}{2} \int dz^2 \;
\left( \;\partial X^\mu  \overline\partial X^\mu
\ - \ \psi_a \overline\partial \psi_a \right)
\cr}
\eqn\hhhhetero
$$
is off-shell supersymmetric.

For scalar fields we need to introduce
$\Phi(z,\overline z , \theta)$
and the selfdual antisymmetric tensor field
${\cal X}^{\mu\nu}(z,\overline z , \theta)$
and demand
$$
{\cal D}_a \sigma^{MN}_{ab} (1) {\cal D}_b \; \Phi \ = \ 0
\ \ \ {\rm and} \ \ \
{\cal D}_a \sigma^{\mu\nu}_{ab} (-1) {\cal D}_b \; \Phi \ =
\ 2 \; \partial {\cal X}^{\mu\nu} \ .
\eqn\hhhact
$$
The latter condition implies that the $\theta^2$-component
of $\Phi$ is the derivative of a scalar.
Now it is not difficult to
verify the supersymmetry of the action
$$
S \ =\ \coeff{1}{8} \int dz^2 \; d\theta_a \;\Big[
 {\cal D}_a \Phi\; \overline\partial \Phi
 \ + \ \coeff{1}{4}
 {\cal D}_a {\cal X}^{\mu\nu}\; \overline\partial {\cal X}^{\mu\nu}
\Big] \ .
\eqn\hhhactone
$$
The introduction of ${\cal X}^{\mu\nu}$ enables us
to introduce four bosonic zeromodes into a theory
that contains a $w=0$ $SO(4)$ scalar.
This trick works of course also for the nonheterotic string.
The result looks component by component exactly
like a vector field, but the interpretation in
terms of $SO(4)$-representations is different.

\section{Ghosts}

The action for ghost-type fields is simpler since
there are no common modes for the left- and rightmoving
fields. We impose the constraints \cconstraints\ and
\bconstraints\ on superfields
${\cal C}(z,\overline z , \theta)$
and ${\cal B}^J(z,\overline z , \theta)$.
Then we may use our construction of the BRST-charge ${\cal Q}$
to immediately write down the off-shell
supersymmetric action
$$
\eqalign{
S\ =& \
- \coeff{1}{12}\int dz^2 \Big[
 \ {\cal D}_a \sigma^J_{ab}(1) {\cal D}_b \ \left(\ {\cal C}
 \overline\partial\;  {\cal B}^J \  \right) \
\cr
& \phantom{- \coeff{1}{12}\int dz^2 \Big[ \ }
 + \coeff{1}{2}
 \left( {\cal D}_a \sigma^J_{ab}(1) {\cal D}_b \; {\cal C} \right)
 \ \overline\partial\; {\cal B}^J
\cr
& \phantom{- \coeff{1}{12}\int dz^2 \Big[ \ }
 - \coeff{1}{2}\ {\cal C}\ {\cal D}_a \sigma^J_{ab}(1) {\cal D}_b
 \; \overline\partial\; {\cal B}^J
 \Big]_{\theta=0} \cr}
\eqn\hhhghostone
$$
The same construction works for all the other
minimal ghosts. We impose, for example,
\cinspinor\ on
${\cal C}_a(z,\overline z , \theta)$
and ${\cal B}^\mu(z,\overline z , \theta)$,
and consequently
$$
S \ = \ -\coeff{1}{4}\int dz^2\; d\theta_a\
\overline\sigma^\mu_{ab}(-1)\;  {\cal C}_b \; \overline\partial\, {\cal B}^\mu
\eqn\hhhghosttwo
$$
is off-shell supersymmetric.
The same is of course true for
$$
S \ = \  -\coeff{1}{4}\int dz^2 \; d\theta_a\
{\cal C}^M \,  \sigma^M_{ab}(1)\;  \overline\partial\, {\cal B}_b
\eqn\hhhghostthree
$$
where the ghost fields obey \cinvector.

The above actions are proper for the heterotic string.
In order to obtain the field content necessary for
closed strings with separate left- and rightmoving
supersymmetries, we simply add in the appropriate
ghost system.
Strictly speaking our ghost actions are in that case not
off-shell supersymmetric.

\chapter{The Lorentz Algebra for the D=6 Superstring}

We will now employ our formalism for analyzing the
light-cone
superstring. Specifically, we will construct the
Lorentz algebra without making explicit reference
to the internal coordinates. Only the internal N=4
superconformal generators $L$, $G_a$ and
$J^{\mu\nu}=\coeff{1}{4}\sigma_{ab}^{\mu\nu}(-1)J_{ab}$
enter the construction.
Their operator products are:
$$
\eqalign{
G_a(z_1) G_b(z_2)\ \ =&\ {4 \delta_{ab}\over z_{12}^3}\ +\
		     {2 \delta_{ab}\over z_{12}} [ L(z_1)+ L(z_2) ]\ +\
		     {1 \over z_{12}^2} [ J_{ab}(z_1) +  J_{ab}(z_2) ] \crr
		   &\ +\ : G_a(z_1) G_b(z_2) : \ ,
\crr
J^{\mu\nu}(z_1)J^{\rho\tau}(z_2)\ =&\
	-\ {4\over z_{12}^2}
	\left[ \delta^{\mu\nu}_{\rho\tau}\ +\ \coeff{1}{2}
	\epsilon^{\mu\nu\rho\tau} \right]
	- {4\over z_{12}} \delta^{[\mu}_{[\rho} \left(
	J^{\nu]}{}_{\tau]}(z_1) + J^{\nu]}{}_{\tau]}(z_2) \right)  \crr
	&\ +\ : J^{\mu\nu}(z_1)J^{\rho\tau}(z_2) : \ ,
\crr
J^{\mu\nu}(z_1) G_a(z_2)\ =&\
	- {1\over z_{12}} \sigma^{\mu\nu}_{ab} G_b(z_2)\
	+\ : J^{\mu\nu}(z_1) G_a(z_2) : \ .
\cr}
\eqn\internalg
$$
The equations above
should also be read as a definition of normal ordered currents.
We emphasize that the $SU(2)$-current $J^{\mu\nu}$ carries
transverse spacetime indices even though it contains no
spacetime coordinates.
The supersymmetry generators are
$$
\eqalign{
Q_a^+ &= 2^{1/4}
	\oint {dz\over 2\pi i} \ \left({ \alpha^+\over z }\right)^{1/2}
 	\psi_a(z)
\cr
Q_a^- &={1\over 2^{1/4} }
	\oint {dz\over 2\pi i} \ \left({z\over \alpha^+ }\right)^{1/2}
 	\left( G_a^x(z) +  G_a(z) \right) \ .
\cr
}
\eqn\susygens
$$
They satisfy the anticommutation relations
$$
\eqalign{
\{ Q_a^+ ,  Q_b^+ \} =& \sqrt{2} \delta_{ab} \alpha^+ \ ;\qquad \qquad
\{ Q_a^- ,  Q_b^+ \} = \sigma^\mu_{ab}
		\oint {dz\over 2\pi i} \partial x^\mu(z)\cr
\{ Q_a^- ,  Q_b^- \} =&  - \sqrt{2}  \delta_{ab}
		\oint {dz\over 2\pi i}  \partial x^-(z)\cr
}
\eqn\susyalgeb
$$
with $\partial x^- = -(\alpha^+)^{-1}z ( L^x + L - \coeff{1}{2}z^{-2})$.
The generators of Lorentz rotations read, at $x_0^+=0$,
$$
\eqalign{
{\bf J}^{+-} &= {1\over 4} \alpha^+ {\partial \over \partial\alpha^+}
\crr
{\bf J}^{+\mu} &= -  {1\over 4} x_0^{\mu} \alpha^+
\crr
{\bf J}^{\mu\nu} &=  \oint {dz\over 2\pi i} \ \Big(
		{1\over 2} x^{[\mu}(z)\partial x^{\nu]}(z)
		+ {1\over 16} \psi(z) \sigma^{\mu\nu}(-1) \psi(z)
		+ {1\over 8} J^{\mu\nu}(z) \Big)
\crr
{\bf J}^{-\mu} &=  \oint {dz\over 2\pi i} \ {z\over \alpha^+}
	\Bigg({1\over 2}x^\mu(z)\Big[ L^x(z) + L(z) - {1\over 2 z^2}\Big]\cr
&\phantom{ \oint {dz\over 2\pi i} \ {z\over \alpha^+}\Bigg(}
            +{1\over 8} \psi \overline\sigma^\mu (-1)
			\Big[G^x(z) + 2 G(z)\Big]
	    +{1\over 4} J^{\mu\nu}(z) \partial x^\nu(z)
	\Bigg) \ .
\cr
}
\eqn\lorentzrot
$$
The $x^\mu$-sector of the algebra is of course identical to
the usual bosonic algebra. Note that the current $\coeff{1}{2}
\psi \sigma^{\mu\nu}(-1) \psi$ appearing in ${\bf J}^{\mu\nu}$
is not the $SU(2)$-current appearing in the N=4 algebra of
the $(x,\psi)$-system.
We have written down only the
open string algebra, or the leftmoving sector of the closed string
algebra. It is fairly obvious how to complete the generators for
the various closed string theories.

The commutator algebra of
the operators defined in \lorentzrot\
yields the desired result if
$c=6$ for the internal algebra and provided
the nullvector condition
$$
: G\sigma^{\mu\nu} G - 4 L J^{\mu\nu} -
J^{\rho[\mu} \partial J^{\nu]\rho}
+ 3 \partial^2 J^{\mu\nu}:\quad = 0\
\eqn\nullvec
$$
is satisfied. More precisely, we need only
require that the zero mode of that operator
vanish.\nullvec\ is related by supersymmetry
transformations to the nullvector
$$
: J^{\mu\nu} J^{\rho\tau} - \coeff{1}{6}
\left( \delta^{\mu\nu}_{\rho\tau} + \coeff{1}{2}
\epsilon^{\mu\nu\rho\tau} \right)
 J^{\lambda\sigma} J^{\lambda\sigma}: \ = 0 \ ,
\eqn\currnull
$$
and its zero mode restricts the allowed
$SU(2)$-representations of the ground
state to carry either spin 0 or spin $1/2$.
An explicit example is furnished by dimensional
reduction of the D=10 superstring:
$$
\eqalign{
L\ \ \  &=\	 \coeff{1}{2} \partial \phi^N \phi^N \
	-\ \coeff{1}{2}\lambda_a\partial\lambda_a
\cr
G_a\  &=\ \sigma^N_{ab}(-1) \lambda_b \partial\phi^N
\cr
J^{\mu\nu} &=\  \coeff{1}{2}\lambda_a\sigma^{\mu\nu}_{ab} (-1)\lambda_b
\cr
}
\eqn\dimredten
$$
One has to pay attention to the fact that there
is a difference between current normal ordered
operators and operators normal ordered with respect
to the modes of $\phi$ and $\lambda$.

It is striking how little room there is for changing
\lorentzrot. Let us assume that the internal sector
appears only through only elements of the N=4 algebra.
We may not actually break $SO(4)$, so we
must have an internal contribution to ${\bf J}^{\mu\nu}$.
Hence an internal N=4 algebra with $SU(2)$-currents
$J^K(z)= \coeff{1}{4}\sigma^K_{ab}(1) J_{ab}$ is impossible.
While $J^x_{ab}$ is selfdual in $[ab]$, $J_{ab}=
\coeff{1}{4}\sigma_{ab}^{\mu\nu}(-1)J^{\mu\nu}$
must be antiselfdual. Higher dimension operators such as
$:G \sigma^{\mu\nu} G: \partial x^\nu$
in ${\bf J}^{-\mu}$ are disallowed since by global scaling they
appear with additional powers of $\alpha^+$, and the
Lorentz algebra requires homogeneity in $\alpha^+$.
One might also try to introduce a dimension 2 primary
vector $A^\mu(z)$ that can be added to ${\bf J}^{-\mu}$
and commutes with itself to the nullvector \nullvec.
This type of mechanism is in fact realized for D=4.
In D=6 such an attempt fails since the superpartner $S^\mu_a(z)$
of $A^\mu(z)$ has to satisfy
$\sigma^{[\mu}_{ab}S^{\nu]}_b =0$, and this implies $S^\mu_a=0$.
We are thus tempted to conjecture that \lorentzrot\ is
the most general form of the Lorentz algebra for the
superstring in flat six-dimensional spacetime.

Lastly, we wish to point out that if we commute ${\bf J}^{-\mu}$
with $\partial x^\mu$ and $\psi_a$, we obtain local operators
that contain the internal N=4 current multiplet.
In this sense the (global) algebra
of \susygens\ and \lorentzrot\ together with the
algebra of spacetime currents $\partial x^\mu$
and $\psi_a$ allows a reconstruction of the local
N=4 current algebra. Essentially, then,
the supersymmetry- and Lorentz-algebra are
an equivalent formulation of that extended
superconformal algebra.

\chapter{Conclusions. What happens for N=8?}

We have developed a superfield formulation for minimal N=4
superconformal field theory in some detail, with emphasis on
the division algebra structure.
We have found that there is a N=4 superconformal symmetry in
the six-dimensional light-cone superstring, and that it is
very directly related to spacetime supersymmetry.
Our superfields are tailormade for this application.

Of course, what we really are after is the
ten-dimensional superstring, and more
specifically its description in a spacetime
covariant superconformal gauge.
Spacetime covariant superstring actions that have a
local extended worldsheet
supersymmetry have been constructed [\confstring,\DGHS],
but it remains to be seen whether they can be gauge
fixed to sufficiently simple covariant
2-D field theories. In particular, the question
of second-class constraints needs to be settled.
The solution to those problems presumably requires
a working knowledge of the appropriate superconformal
field theory, and we have provided the necessary
machinery for D=6.

We have not yet completed the superfield formulation
of N=8 superconformal field theory. The construction
principles are clear however, and we will describe them
in words.
The N=8 model carries much of the structure of
the octonion algebra. That algebra is nonassociative, and
apart from that fact the generalization from N=4 to N=8
is straightforward.
A general N=8 scalar superfield may be constrained
analogously to eq.\dualityconstr, and
the component fields consist of eight bosons and
eight fermions, transforming in the desired way. The
projection operators analogous to \projectop\ contain instead
of the $\varepsilon$-symbol a tensor antisymmetric in four
spinor indices [\dewitetal], constructed from the octonionic
structure constants and projecting onto 7- and 21-dimensional
subspaces in the antisymmetric product.

The analogue of $SU(2)\approx S^3$ for N=4 is
$S^7$, the round seven-sphere. Of course, $S^7$ is not
a group manifold anymore, and that has to be taken
properly into account. The quaternionic structure
constants have to be regarded as components of the
parallelizing torsion tensor, and consequently the
gamma-matrices also become tensors on $S^7$ [\rooman,\ESTPS].
The associator terms that spoil the naive generalization
from N=4 to N=8 are then interpreted as
covariant derivatives of the torsion tensor on the sphere.
We emphasize that the superconformal algebra we construct
is  neither of the field-dependent type [\ESTPS,\DTH,\Berkovits]
nor of the exceptional type [\FLBow].

The relationship between the light-cone super-Poincar\'e
algebra and the worldsheet superconformal algebra that
we discovered for N=4 carries over to the N=8 construction
outlined above.

A more comprehensive analysis of octonionic conformal field
theory will be forthcoming, as well as an expanded discussion
of the relation between light-cone super-Poincar\'e algebras
and N=2,4,8 superconformal field theory.

\ack We would like to thank Lars Brink and Bengt E.W. Nilsson
for many long and interesting discussions, and for taking
active part during the initiation process of this work.
Many of our ideas about the light-cone superstring and
octonionic conformal field theory were developed in
conversations with Lars Brink.

\refout

\endpage

\Appendix{}

We define
$$
\sigma^0_{ab}(e) = \delta_{ab} \ \ \ \ {\rm and} \ \ \ \
\sigma^j_{ab}(e) = -2e \delta^0_{[a}\delta^j_{b]} - \epsilon_{j\alpha\beta}
\ ,\eqno\eq
$$
where $a=(0,\alpha)$, $e=\pm 1$ and $[ab]=\coeff{1}{2} (ab - ba)$.
Note the selfduality property
$$
\sigma^j_{ab}(e) =
\coeff{1}{2}(\delta^{ab}_{cd}+ \coeff{e}{2}\varepsilon_{abcd})
\sigma^j_{cd}(e) \ ,\eqno\eq
$$
where $\delta^{ab}_{cd}= \delta^{[a}_c \delta^{b]}_d$.
For $\sigma^\mu = (\sigma^0,\sigma^j)$ and
$\overline{\sigma}^\mu =  (\sigma^0,-\sigma^j)$
the chiral Lorentz generators are (anti)selfdual:
$$
\eqalign{
\sigma^{[\mu}(e)\overline{\sigma}^{\nu]}(e) \equiv\sigma^{\mu\nu}(e)
&= \coeff{1}{2}\left[ \delta^{\mu\nu}_{\rho\tau} + \coeff{1}{2}
\epsilon^{\mu\nu\rho\tau} \right]  \sigma^{\rho\tau}(e)
 \cr
\overline\sigma^{[\mu}(e){\sigma}^{\nu]}(e) \equiv\overline\sigma^{\mu\nu}(e)
&= \coeff{1}{2}\left[ \delta^{\mu\nu}_{\rho\tau} - \coeff{1}{2}
\epsilon^{\mu\nu\rho\tau} \right]  \overline\sigma^{\rho\tau}(e)
 \cr}
\eqno\eq
$$
The generators of the four different $SU(2)$'s are obtained by
(anti)selfduality projections on the spinor indices, i.e. by
choices of $e=\pm 1$.
We set $e^\mu = (1,e^j)$ as well as $e_a = (1,e_\alpha)$ with the
quaternion units $e_j$ satisfying the same algebra as the $\sigma^j$:
$$
e^i e^j = - \delta^{ij} + \epsilon^{ijk} e^k \ .\eqno\eq
$$
Then the following identities hold:
$$
\eqalign{
e^*_a e_b =& \sigma^\mu_{ab}(-1) e^\mu =
             \overline{\sigma}^\mu_{ab}(-1) e^{\mu *} \cr
e_a e_b^* =& \sigma^\mu_{ab}(1) e^\mu\ \ =
             \overline{\sigma}^\mu_{ab}(1) e^{\mu *} \cr}
\eqno\eq
$$
$$
\eqalign{
e_a e^\mu =& \sigma^\mu_{ab}(-1) e_b \ \ \ \
e_a e^{\mu *} =\overline \sigma^\mu_{ab}(-1) e_b
\cr
e_a^* e^\mu =& \sigma^\mu_{ab}(1) e_b^* \ \ \ \ \ \
e_a^* e^{\mu *} =\overline \sigma^\mu_{ab}(1) e_b^*
\cr}
\eqno\eq
$$
$$
\eqalign{
e^\mu e^{\nu *} =& \delta^{\mu\nu} +
\coeff{1}{2}\left[ \delta^{\mu\nu}_{\rho\tau} + \coeff{1}{2}
\epsilon^{\mu\nu\rho\tau} \right] e^\rho e^{\tau *}
=
 \delta^{\mu\nu}
-2\left[ \delta^{\mu\nu}_{0j} + \coeff{1}{2}
\epsilon^{\mu\nu 0 j} \right] e^j
 \cr
e^{\mu *} e^{\nu } =& \delta^{\mu\nu} -
\coeff{1}{2}\left[ \delta^{\mu\nu}_{\rho\tau} - \coeff{1}{2}
\epsilon^{\mu\nu\rho\tau} \right] e^{\rho *} e^{\tau }
=
 \delta^{\mu\nu}
+2\left[ \delta^{\mu\nu}_{0j} - \coeff{1}{2}
\epsilon^{\mu\nu 0 j} \right] e^j
\cr}
\eqno\eq
$$
Note that $e^\mu e_a = (e_a^* e^{\mu *})^*$.
The above equations provide a handy translation table for conversion from
quaternion notation to $\sigma$-matrix language.

\end